**Facile size-controllable synthesis of single crystalline *β* -MnO$_2$ nanorods under varying acidic strengths**

*Niraj Kumar[a], P. Dineshkumar[a], R. Rameshbabu[a], and Arijit Sen[a,b,*]*

[a]Department of Physics and Nanotechnology, SRM University, Kattankulathur-603203, India

[b]SRM Research Institute, SRM University, Kattankulathur-603203, India

[*]Corresponding author E-mail: arijit.s@res.srmuniv.ac.in

**Abstract**

A simple one-pot hydrothermal synthesis of single crystalline *β*-MnO$_2$ nanorods with diameters in the range of 10-40 nm is reported. During the synthesis process, the acid molarities were varied from 1.1 M down to 0.2 M in steps of 0.3 M while keeping the other reaction parameters constant, resulting in gradual transformation of the size of *β*-MnO$_2$ from micro to the nanoscale dimension. The as synthesized nanorods exhibit soft ferromagnetic behavior and possess a high catalytic activity with an onset potential of –0.17 V in facilitating the oxygen reduction reaction (ORR).

**1. Introduction**

Single crystalline nanorods and nanowires have garnered immense enthusiasm in recent years because of their foreseeable applications as interconnects and building blocks in various nanoelectronic devices [1-3]. Quantum effects related to the shape and size of such nanostructures are considered to influence their physical properties, which depart considerably from those of the respective bulk phases, and in turn, hold a key factor to the ultimate performance and application of nanoscale materials [4]. In view of this, low dimensional manganese dioxide (MnO$_2$) promises to exhibit fast electro-kinetics and appreciably moderate catalytic activity, due mainly to its high surface area and also, large number of active sites. Besides, for its redox stability, catalytic activity and environmentally benign nature along with reduced toxicity, earth-abundance, and low cost, nano-structured MnO$_2$ often poses a natural selection in various technological applications such as ion-based batteries [5], super-capacitors [6-9], magnetic nanomaterials [10], and fuel-cell catalysts [11-14].

Although platinum (Pt) based catalysts are commonly used in fuel cells to enhance the performance of oxygen reduction reaction (ORR) with remarkable efficiency, its limited availability hinder the scaling up in a cost-effective way. Hence, considerable research efforts have been in place for quite some time to explore non-precious catalysts such as manganese oxides as possible alternative to Pt in fuel cells. Recently, Zhang et al. [15] reported that $\beta$-MnO$_2$ have the potential to improve the feasibility of scaling-up microbial fuel cells (MFCs) for realistic applications. In another work, Liu et al. [16] illustrated that nano-structured MnO$_x$, as obtained through an electrochemical deposition method, is an effective ORR catalyst in MFCs.

MnO$_2$ is known to exist in many polymorphic forms including $\alpha$, $\beta$ and $\gamma$ type according to different linking manners of the basic unit [MnO$_6$] octahedral structure [17]. Numerous techniques have been developed for the synthesis of nanoscale manganese dioxides including sol-gel templating [18], thermal decomposition [19], refluxing [20], solvent free solid reaction [21], electrodeposition [22] and hydrothermal methods [23-28]. Recently, we have demonstrated a simpler and low-cost hydrothermal method for the synthesis of ultrafine $\alpha$-MnO$_2$ nanowires [29]. In the present work, we adopt quite a similar approach to examine the morphological growth of single crystalline $\beta$-MnO$_2$ nanorods under, however, different acidic conditions. From the best of our knowledge, few works have been done to systematically analyze the morphological growth of $\beta$-MnO$_2$ nanorods under varying acidic strengths. The ORR activity and magnetic hysteresis of such nanorods are also studied to evaluate their electrochemical and magnetic properties.

## 2. Experimental

### 2.1 Materials

Potassium permanganate (KMnO$_4$), Sodium nitrite (NaNO$_2$), Sulfuric acid (H$_2$SO$_4$), Potassium hydroxide (KOH), Nafion solution (C$_7$HF$_{13}$O$_5$S·C$_2$F$_4$; 5 wt % in water) and Isopropyl alcohol (C$_3$H$_8$O) were purchased from Sigma Aldrich. The reagents used were of analytical grade with high purity and were used without any further treatment.

### 2.2 Synthesis

In a typical synthesis, 8 mM KMnO$_4$ and 12 mM NaNO$_2$ (2:3 molar ratio) were mixed in 35 ml of de-ionized water. Then, 0.2 M H$_2$SO$_4$ was prepared in 5 ml of de-ionized water and this solution was added drop wise into the above solution under continuous stirring to form a homogeneous solution of 40 ml. The prepared solution was bolted under airtight condition inside a teflon-lined stainless steel autoclave (50 ml) of 80% capacity of the total volume. The autoclave was kept in muffle furnace and the hydrothermal process was carried out at a temperature of 170 $^o$C for 12 h. Then, the autoclave was cooled to room temperature naturally. Next, the precipitates were centrifuged and thoroughly washed with de-ionized water and ethanol for several times and were dried in hot air oven at 100 $^o$C for 5 h to obtain the final product. The final product obtained was calcined at 400 $^o$C for 6 h and was named as sample S4. Meanwhile, for size analysis a series of control experiments were performed under varying acidic strengths by inceasing the molarity of $H_2SO_4$ from 0.2 M to 1.1 M in steps of 0.3 M while keeping the other synthesis parameters constant. And the subsequent final products were named as sample S3, S2 and S1 respectively.

**2.3 Electrode preparation**

The catalyst slurry was prepared by mixing 9 mg of sample S1 with 5 mL Isopropyl alcohol (24%) and 25 µL Nafion solution (5 wt %). The mixure was then ultrasonicated for 30 min in ice cold water to form homogenous slurry. Then, 4 µL of the as prepared slurry was spun onto the surface of 3 mm glassy carbon (GC) electrode. The electrode coated with the catalyst was then dried at room temperature for 6 h and was used as a working electrode. The same process was carried out with sample S2, S3 and S4.

**2.4 Characterization**

PAN analytical X' Pert Pro diffractometer was used for the powder X-ray diffraction (XRD) measurement employing Cu-Kα rays of wavelength 1.5406 Å with a tube current of 30 mA at 40 kV in the 2 θ range of 10-80 degree. The morphological analyses of the samples were examined by Field emission scanning electron microscopy (Quanta 200 FEG FE-SEM). Energy dispersive X-ray analysis (EDX) was done using 'Bruker 129 ev' with 'Espirit software'. 'HR-TEM, JEM-2010, 200kV' was also used for High resolution Transmission electron microscopy for morphological and energy dispersive X-ray analysis (EDX) of $MnO_2$ nanostructures. The specimen for TEM analysis was prepared by suspending powder samples in acetone. Fourier transform infrared spectroscopy (FTIR) using Perkin Elmer Spectrophotometer was used for the functional group analysis of the material by the KBr pellet technique in the range of 400-2000 $cm^{-1}$. The magnetization curves of the materials were measured by a VSM lakeshore-7410 at room temperature.

Cyclic voltammetry (CV) measurements were conducted using a CHI604E potentiostat with a three electrode system. A 3 mm diameter glassy carbon electrode coated with the catalyst, a platinum (Pt) wire, and silver/silver chloride (Ag/AgCl, [saturated KCl], 222 mV vs. standard hydrogen electrode) were used as the working, counter and reference electrodes, respectively. The CV measurements were performed from –0.9 to 0.1 V at a scan rate of 100 mV/s in an aqueous electrolyte system (0.1 M KOH). Before each scan series, the electrolyte solution was bubbled with $O_2$ (or $N_2$) for 30 min to establish an aerobic (or anaerobic) environment. Furthermore, the electrochemical oxygen reduction reaction (ORR) activities of the *β* -$MnO_2$ nanorods were studied using linear sweep voltammetry in a 0.1 M KOH aqueous electrolyte with a rotating disk electrode at different rotation rates of 400, 900, 1600 and 2500 rpm and scan rate of 10 mV/s.

Surface area and porosity measurements were carried out using a Quantachrome Nova-1000 surface analyzer at liquid nitrogen temperature.

### 3. Results and Discussions

### 3.1 Morphological Analysis

The redox reaction between permanganate and nitrite ions with proton involvement from the sulphuric acid is the main principle behind the production of $\beta$-MnO$_2$ nanorods, which described as follows [30]:

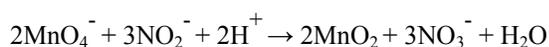

$$2MnO_4^- + 3NO_2^- + 2H^+ \rightarrow 2MnO_2 + 3NO_3^- + H_2O$$

Here, permanganate ion acts as an oxidant and resource of manganese (Mn) element, whereas nitrite ions are used as a reducing agent. From the equation it can be observed that the stoichiometric molar ratio of permanganate to nitrite ions is 2:3. Le Chatelier's principle suggested the involvement of proton in this redox reaction. Optimizing the amount of proton helps in the formation of well aligned nanorods.

Figure 1(a-d) shows the FESEM images of samples S1-S4 respectively prepared under diffrent acidic conditions. The size of the morphologies in the samples S1-S4 were calculated from the average values based on the measurements of 100 similar kind of structures, respectively using the image software pre-installed in "Quanta 200 FEG FE-SEM instrument". Sample S1 has the rods like structures in micro as well as submicro regime, having diameters in range of 0.4-2 μm. Sample S2 has the submicrorods like structures with lesser diameters of 0.2-1 μm. This is due to the lesser amount of proton involvement (H$_2$SO$_4$) during the synthesis of sample S2 i.e. 0.8 M H$_2$SO$_4$ for S2 compared to 1.1 M H$_2$SO$_4$ for S1. Decrease in amount of protons tends to increase the reduction potential of permangante ions based on Nernst equation [31]. This increase in reduction potential can be compensated by a decrease in reduction potential of nitrite ions for equilibrium conditions. This restricts the redox reaction between permanganate and nitrite ions resulting in the formation of smaller dimensional MnO$_2$ structures. This is evident from the further decrease in amount of H$_2$SO$_4$ from 0.8 M to 0.5 M and 0.2 M, lead to the transformations in the sample from micro to nano dimensions. Sample S3 prepared with 0.5 M exhibits nanowires like structures with diameters in the range of 10-40 nm, whereas sample S4 prepared with 0.2 M represents nanorods like structure having similar diameters but smaller lengths (0.4-1 μm) than sample S3. However, the length difference between samples S3, S2 and S1 is not visualized properly. Furthermore, it can be predicted that only under optimized conditions with relatively less acidic effect, the probabilities of getting nanostructures are high.

### 3.2 Structural Analysis

Fig. 1(e) shows the XRD patterns of the samples S1-S4. The high intensity major diffraction peaks at 2θ = 28.68, 37.33 and 56.65 were observed for all the samples. The relative order of intensities of various peaks for all the respective samples can be assigned to a pure crystalline tetragonal phase of β-$MnO_2$ (JCPDS 24-0735). Further, no peaks for amorphous or other types of $MnO_2$ were observed in the XRD spectra, suggesting high purity and crystallinity of the final products. The XRD peak of (110) plane gradually becomes stronger on decreasing the acid ($H_2SO_4$) concentration from 1.1 M to 0.2 M in steps of 0.3 M for samples S1-S4, respectively. This indicates that the subsequent formation of β-$MnO_2$ crystals preferably grow along the (110) plane direction. The intensities as well as broadening of the corresponding peaks follow an increasing order from sample S1 to S4 respectively. Sample S4 has the highest and broadest peaks, suggesting the formation of single crystalline β-$MnO_2$ nanostructures. The high broad peaks of S3 also depict the presence of nanostructures, whereas the less intense and narrower peaks observed for samples S2 suggests the presence of submicrostructures. Sample S1 has the lowest and narrowest peaks for their large micro dimensional shape compared with the other samples. The results are found in consistent with the FESEM analysis.

A schematic representation on the morphological changes of the as prepared samples S1-S4 under different acidic conditions of 1.1 M, 0.8 M, 0.5 M and 0.2 M $H_2SO_4$ respectively, is illustrated in Figure 2. It clearly portrays that the dimensions of the final products (S1-S4) decreases on decreasing the acid concentration. Sample S4 is represented as the single crystalline β-$MnO_2$ nanorods, whereas samples S3, S2 and S1 are represented by nanowires, submicrorods and a mixture of micro as well as submicrorods, respectively.

### 3.3 Microstructure Analysis

The TEM and HRTEM images shown in figures 2 (a-d) revealed that the β-$MnO_2$ nanorods (sample S4) obtained with 0.2 M $H_2SO_4$ exhibits ultrafine morphology. This further confirms that nanorods obtained had one dimensional structure with diameter in the range of 10-40 nm. Figure 3(e) shows the selected area electron diffraction pattern of the corresponding sample which supports the formation of single crystalline tetragonal structure of β-$MnO_2$ nanorods. Energy dispersive X-ray analysis (EDX) of the corresponding sample is represented in figure 3(f). It clearly shows the high intensity peaks of Manganese and Oxygen elements and hence indicates the obtained nanorods were in pure form with negligible impurities.

### 3.4 Functional group Analysis

Functional group analysis of all the samples was performed using the FTIR spectroscopy as shown in Fig. 4(a). The bands at about 715, 531 and 448 $cm^{-1}$ that are below 750 $cm^{-1}$ can be attributed to the metal-oxygen (Mn–O) bending modes of $\beta$-$MnO_2$. Similar FTIR patterns were observed for all the samples (S1-S4). It suggests a high structural symmetry of $\beta$-$MnO_2$ in all the samples. The functional group analysis carried out here is found to be in good agreement with the results reported in the literature [32], and is also consistent with the XRD data.

### 3.5 Magnetic Properties

Magnetic properties of the as-synthesized products were investigated using a vibrating sample magnetometer with an applied field of ±12500 Oe at room temperature. The plot of magnetization vs. applied magnetic field for the samples (S1-S4) is given in Fig. 4(b). At low external field, the hysteresis loop exhibits low coercivities, which is characteristic of soft ferromagnetism. The lowest coercivity ($H_c$ = 110.8 Oe) was observed for the sample S4, suggesting that it has smaller dimension (*i.e.* nanoscale) than all other samples. The coercivity increases with the increasing dimension of the material. This is well illustrated, as the samples S2 and S3 showcase higher coercivity values of 111.5 and 112.2 Oe respectively with respect to the sample S4. The highest coercivity value of 112.5 Oe was observed with the sample S1 due its microscale dimension. The magnetic domains in nanoscale materials are less as compared to their bulk counterparts, and so they can be easily demagnetized showing low coercivity and high saturation magnetization. At high external field, the dependence of magnetization on applied field is markedly linear and saturation magnetization is not reached in any of the samples even at the maximum applied field of 12500 Oe. However, an increasing trend in the saturation magnetization (Ms) value for sample S1-S4 can be predicted from their peak magnetization values of 0.040, 0.045, 0.067 and 0.083 emu/g, respectively, as seen from figure 4(b). Since the coercivity is directly proportional while the saturation magnetization is inversely proportional to the size of the material, it can be assumed that there was a size reduction in the samples S1-S4 from micro to nanostructures. This assumption goes well with our experimental observation.

The magnetic properties are dependent on the valency of Mn in the $[Mn_2]O_4$ compound. Thus the magnetic properties are evaluated by interactions between the Mn ions only [33]. The magnetic response of $\beta$-$MnO_2$ could be ascribed to the ferromagnetic coupling of $Mn^{4+}-O^{2-}-Mn^{4+}$, whereas all other interactions are in antiferromagnetic coupling.

### 3.6 Electrochemical Properties

Figure 5(a) shows the cyclic voltammograms of electrocatalysts S1-S4 for oxygen-reduction-reaction (ORR) at the scan rate of 100 mV/s in 0.1 M KOH. It can be observed that the hydrothermally synthesized samples S1-S4 give reduction peaks of –0.51, –0.49, –0.47 and –0.42 V, respectively, in the aerobic environment ($O_2$ saturation). These peaks affirm the catalyzed ORR activity of $β$-$MnO_2$. An increasing trend in the reduction peaks is accounted for the decrease in size of the catalyst. As the surface energy increases with the decrease in the catalyst size, the sample S4, having the smallest size, exhibits the highest reduction peak, portraying a higher ORR activity. Interestingly, these synthesized catalysts display similar electrochemical properties, as reported in the works of Xiao et al. [34] and Yuan et al. [35].

To get insights into the oxygen reduction kinetics of the catalysts, rotating ring disk electrode (RDE) measurements were performed. Figure 5(b) shows the RDE voltammograms in $O_2$ saturated 0.1 M KOH of the catalysts. An analogous trend in the onset potential and ORR current densities of the electrocatalysts was observed. From figure 5(b), the increasing order of onset potentials for catalysts S1-S4 are respectively –0.30, –0.28, –0.18 and –0.17 V vs. Ag/AgCl, and ORR current densities are respectively 3.14, 3.45, 4.06 and 5.04 mA/cm$^2$ at –0.9 V. High onset potential as well as current density as yielded by S4 is obvious because of its smaller size as compared to other samples.

A detailed analysis on this increasing trend of onset potentials and current densities with decreasing catalyst size was performed for the samples S1-S4 at different rotating disk speeds. At –0.9 V, S1 yields ORR current densities of 1.46, 1.96, 2.47 and 3.14 mA/cm$^2$ for different rotating disk speeds of 400, 900, 1600 and 2500 rpm, respectively [figure 6(a)]. Similarly, S2 yields ORR current densities of 1.62, 2.17, 2.71 and 3.45 mA/cm$^2$ [figure 6(b)], S3 yields ORR current densities of 1.97, 2.65, 3.24 and 4.06 mA/cm$^2$ [figure 6(c)] and S4 yields ORR current densities of 2.44, 3.29, 4.02 and 5.04 mA/cm$^2$ for different rotating disk speeds of 400, 900, 1600 and 2500 rpm, respectively [figure 6(d)]. This analysis is in corollary with the above results.

The positive shift of the onset potential and the enhancement in the reduction current densities compared to the as-prepared $MnO_2$ nanostructures by H. Yuan et al. [35] (which had the onset potential of –0.18 V) for the ORR clearly indicates that the catalyst S4 possesses improved catalytic activities. It may be due to the fact that reduction of diameters in nanoscale regime of as-prepared nanorods results in the formation of highly confined regions with high surface energies. This increases the tension between the interatomic layers in nanorods generating increased number of active sites, which, in turn, enables the adsorption of more $O_2$ molecules inside these layers, and results in the further decrease of the ORR over-potential. As a result of this, porosity of the S4 (Table 1) gets enhanced, which might shorten the ion transport distances and provide a continuous pathway for the rapid diffusion of electrolytes, thereby improving the electrochemical performance of the electrode materials.

### 3.7 Surface Analysis

The BET surface areas, pore volumes, and pore diameters are indexed in table 1.

**Table 1**

BET surface areas, pore diameter and pore volume

| Sample | Surface area ($m^2/g$) | Average pore diameter (nm) | Total pore volume ($cm^3/g$) |
|---|---|---|---|
| S1 | 09.783 | 18.517 | 0.045 |
| S2 | 14.526 | 14.344 | 0.050 |
| S3 | 30.644 | 09.259 | 0.071 |
| S4 | 37.900 | 07.600 | 0.072 |

BET surface areas of $β$-$MnO_2$ for samples S1-S4 follows an increasing order with S1 having the minimum surface area (9.783 $m^2/g$) and S4 having the maximum surface area (37.9 $m^2/g$) along with S2 and S3 as the intermediates. In addition, the average pore diameter is reduced in the following order: S1>S2>S3>S4. Scientifically, the surface area increases with decrease in size of the material, in addition, there is a decrease in average pore diameter. This supports that there were size transformations in samples S1 to S4 from higher to lower dimensions during the syntheses with S1 and S4 having the highest and lowest dimensional structures.

Further, the total pore volume for sample S4 (0.072 $cm^3/g$) was maximum and sample S1 (0.045 $cm^3/g$) was minimum. A relatively low pore volume as compared with the reported data [36] confirms the high rigidity and nonporosity nature of the as-synthesized $β$-$MnO_2$. However, a higher pore volume observed for the sample S4 as compared with the other samples adds to its high electrochemical activity as demonstrated in the above results.

**4. Conclusions**

Single crystalline $β$-$MnO_2$ nanorods were successfully synthesized under optimized conditions by a simple hydrothermal method without using any template, seed or other capping agent. The influence of different acidic strengths on the morphology of $β$-$MnO_2$ during its growth process was studied systematically. It turns out that the varying molarities of $H_2SO_4$ during the hydrothermal syntheses, can alone lead to the size transformations from micro to nanoscale. The dependency of magnetic and electrochemical properties on the size of the material were evaluated successfully to demonstrate the size reduction which was further held by the surface analyses of the as-obtained samples.

**5. Acknowledgements**

One of the authors, Niraj Kumar, is grateful for the financial support from SRM University and also, thankful to the Department of Physics & Nanotechnology for XRD and FTIR analysis; NRC for FESEM measurements; SRM Research Institute for electrochemical analysis; NCNSNT, University of Madras for HRTEM measurements, and Pondicheery University for VSM measurements.

**Figure Captions**

Fig. 1.  FESEM images of the samples S1-S4, prepared using $KMnO_4$ and $NaNO_2$ in the molar ratio of 2:3 at a temperature of 170 $^oC$ for 12 h with $H_2SO_4$ in the molar concentration of (a) 1.1 M, (b) 0.8 M, (c) 0.5 M and (d) 0.2 M respectively; (e) XRD patterns of the samples S1-S4.

Fig. 2.  Schematic diagram representing morphological changes in the *β*-$MnO_2$ structure under varying acidic strengths.

Fig. 3.  (a-d) TEM and HRTEM images at different magnifications, (e) SAED pattern and (f) EDX analysis of the sample S4, prepared using $KMnO_4$ and $NaNO_2$ in the molar ratio of 2:3 at a temperature of 170 $^oC$ for 12 h with 0.2 M $H_2SO_4$.

Fig. 4.  (a) FTIR spectra and (b) room temperature magnetic property measurements of the samples S1-S4, prepared using $KMnO_4$ and $NaNO_2$ in the molar ratio of 2:3 at a temperature of 170 $^oC$ for 12 h with $H_2SO_4$ in the molar concentration of 1.1 M, 0.8 M, 0.5 M and 0.2 M, respectively.

Fig. 5.  (a) Cyclic voltammograms for ORR at a scan rate of 100 mV/s in 0.1 M KOH under $O_2$ saturations and (b) Rotating disk voltammograms (RDE) in $O_2$ saturated 0.1 M KOH at 2500 rpm for samples S1-S4, prepared using $KMnO_4$ and $NaNO_2$ in the molar ratio of 2:3 at a temperature of 170 $^oC$ for 12 h with $H_2SO_4$ in the molar concentration of 1.1 M, 0.8 M, 0.5 M and 0.2 M, respectively.

Fig. 6.  Rotating disk voltammograms in $O_2$ saturated 0.1 M KOH at 400, 900, 1600 and 2500 rpm for the samples, prepared using $KMnO_4$ and $NaNO_2$ in the molar ratio of 2:3 at a temperature of 170 $^oC$ for 12 h with $H_2SO_4$ in the molar concentration of 1.1 M, (a) sample S1, 0.8 M, (b) sample S2, 0.5 M, (c) sample S3 and 0.2 M, (d) sample S4, respectively.

**Figures**

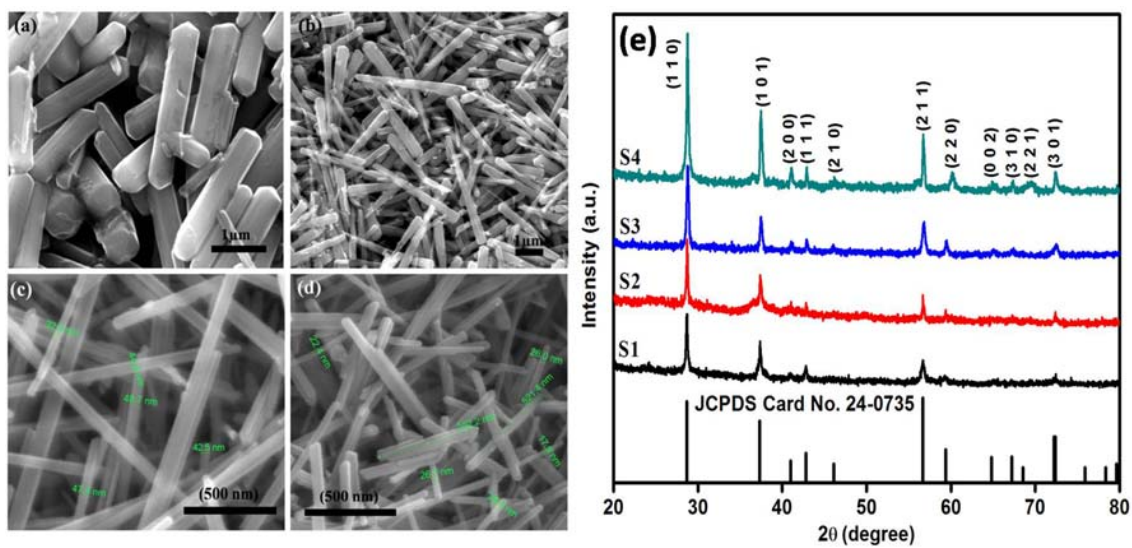

**Figure 1**

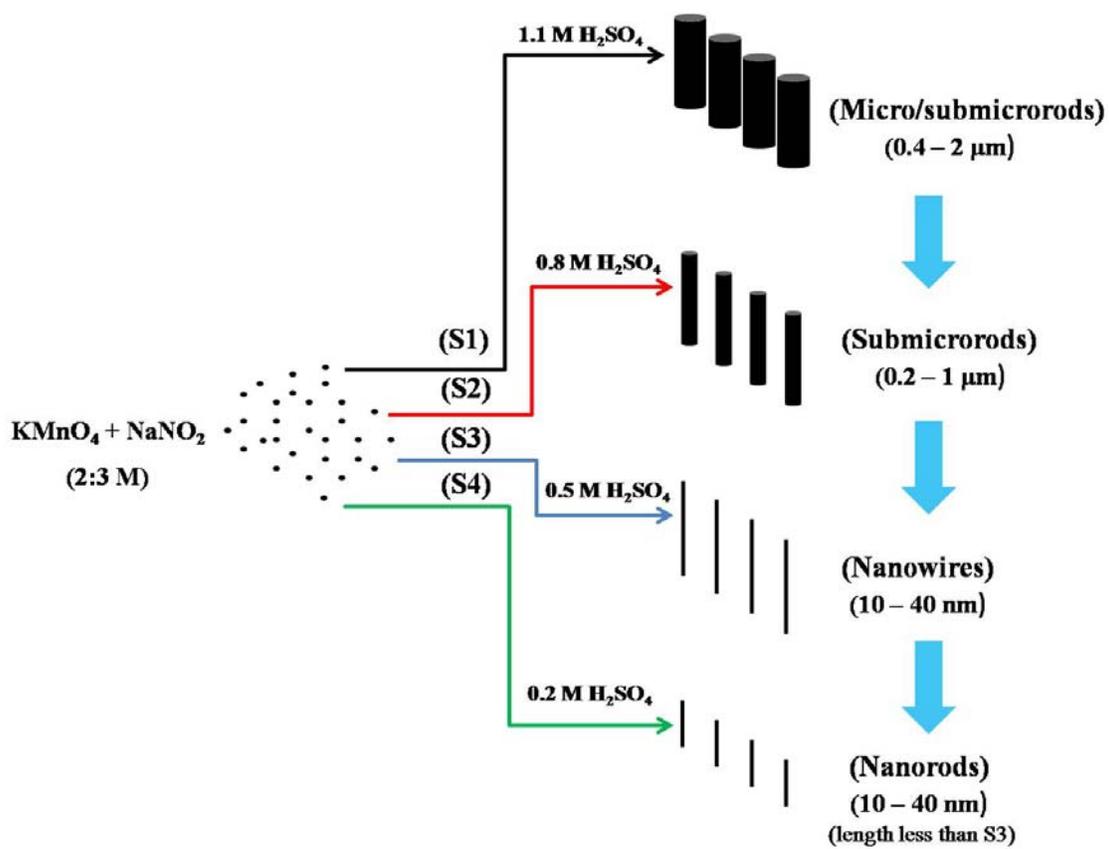

**Figure 2**

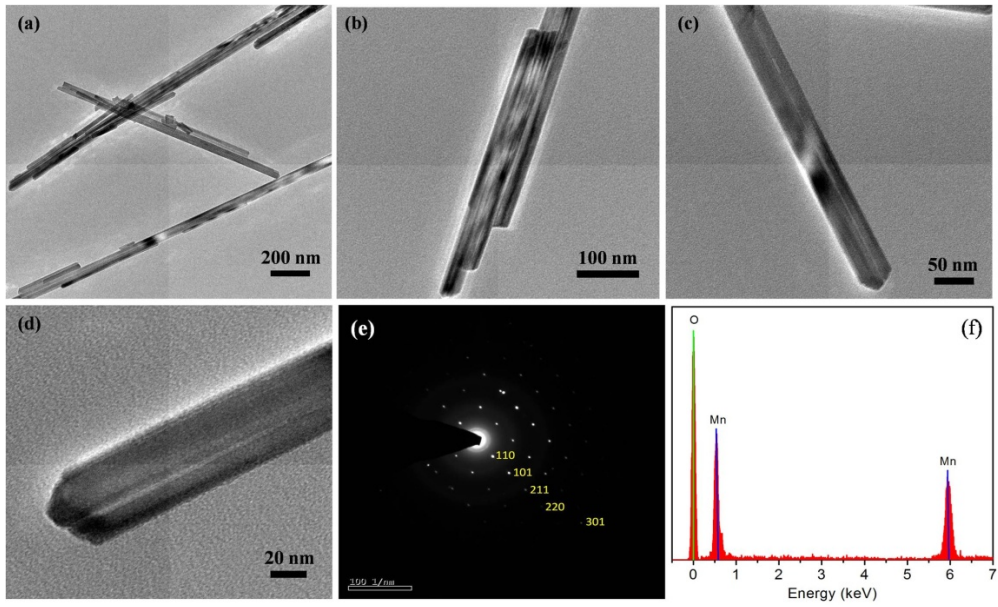

**Figure 3**

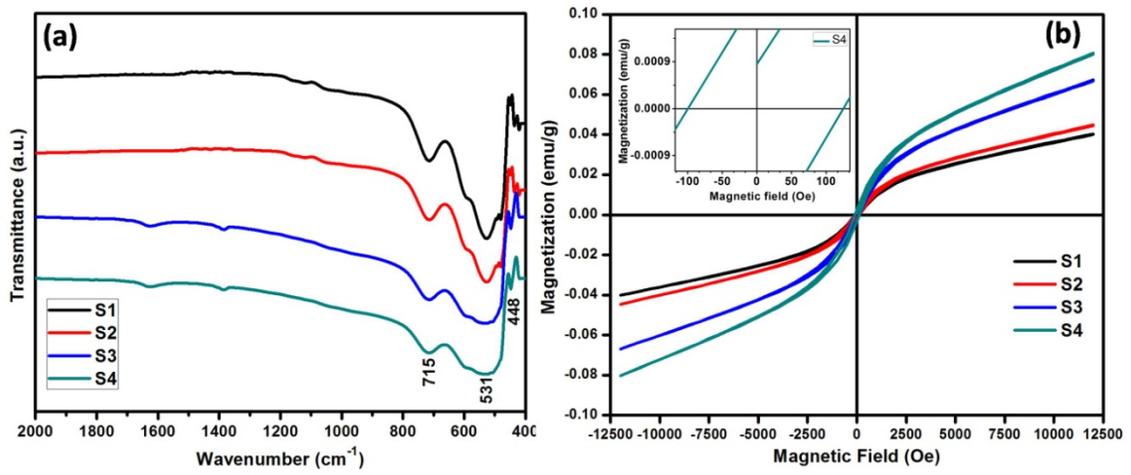

**Figure 4**

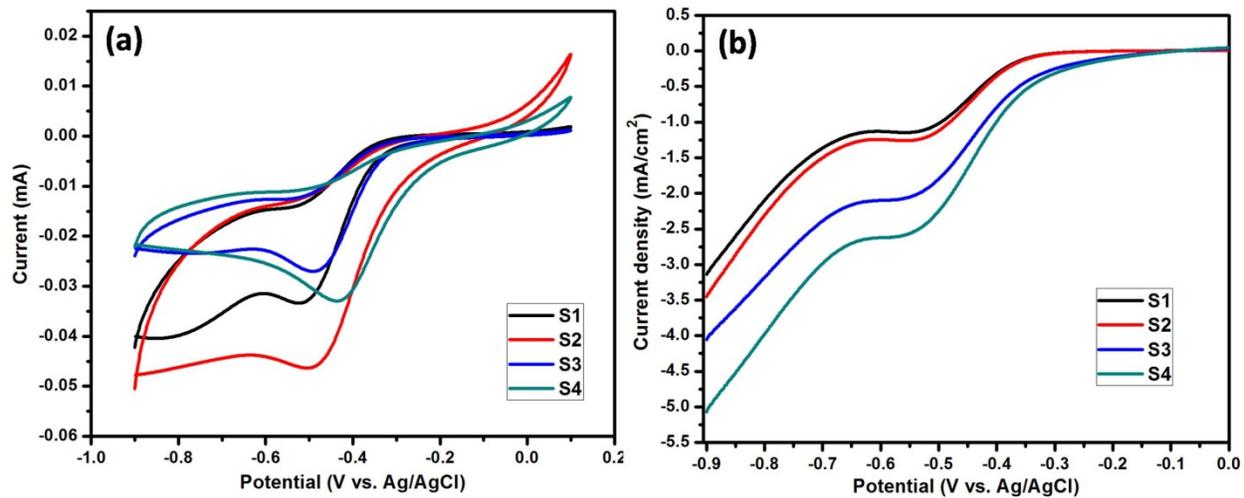

**Figure 5**

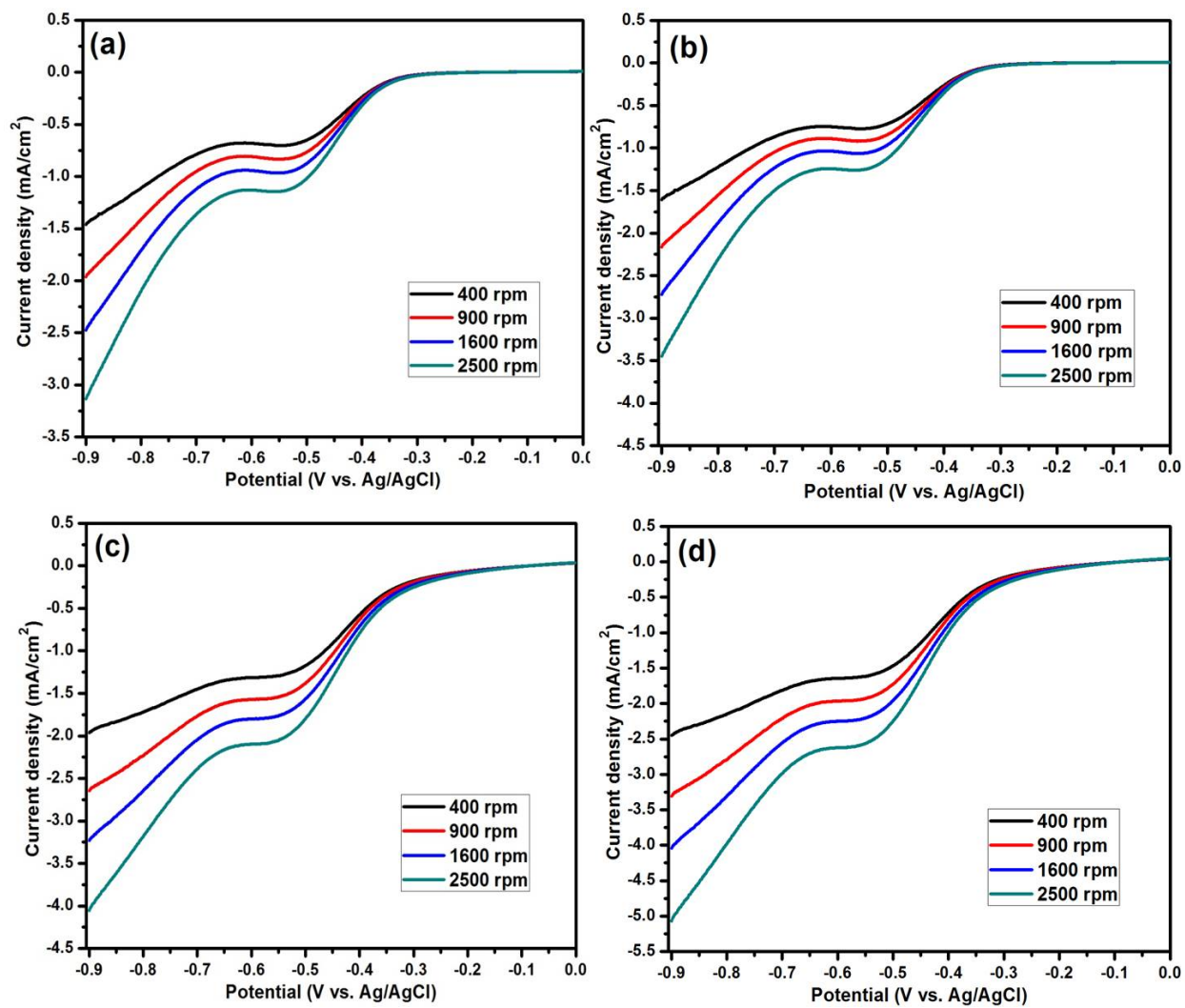

**Figure 6**